\documentclass[prl,showpacs,twocolumn,superscriptaddress]{revtex4}
\usepackage[dvips]{graphicx}
\usepackage{amssymb}
\newcommand{\ped}[1]{\ensuremath{_{\rm #1}}}

\begin{document}

\title{Evidence for Single-gap Superconductivity in Mg(B$_{1-x}$C$_x$)$_2$
Single Crystals with $x=0.132$ from Point-Contact Spectroscopy}

\author{R.\@S. Gonnelli}
\affiliation{Dipartimento di Fisica and INFM, Politecnico di Torino,
10129 Torino, Italy}
\affiliation{INFM - LAMIA, Corso Perrone 24, 16152 Genova, Italy }%
\author{D. Daghero}
\affiliation{Dipartimento di Fisica and INFM, Politecnico di Torino,
10129 Torino, Italy}
\affiliation{INFM - LAMIA, Corso Perrone 24, 16152 Genova, Italy }%
\author{A. Calzolari}
\affiliation{Dipartimento di Fisica and INFM, Politecnico di Torino,
10129 Torino, Italy}
\author{G.\@A. Ummarino}
\affiliation{Dipartimento di Fisica and INFM, Politecnico di Torino,
10129 Torino, Italy}
\affiliation{INFM - LAMIA, Corso Perrone 24, 16152 Genova, Italy }%
\author{Valeria~Dellarocca}
\affiliation{Dipartimento di Fisica and INFM, Politecnico di Torino,
10129 Torino, Italy}
\author{V.\@A. Stepanov}
\affiliation{P.N. Lebedev Physical Institute, Russian Academy of
Sciences, 119991 Moscow, Russia}
\author{S.\@M. Kazakov}
\affiliation{Solid State Physics Laboratory, ETH, CH-8093 Zurich,
Switzerland}
\author{N. Zhigadlo}
\affiliation{Solid State Physics Laboratory, ETH, CH-8093 Zurich,
Switzerland}
\author{J. Karpinski}
\affiliation{Solid State Physics Laboratory, ETH, CH-8093 Zurich,
Switzerland}


\begin{abstract}
We report the results of the first directional point-contact
measurements in Mg(B$\ped{1-x}$C$\ped{x}$)$_2$ single crystals
with $0.047 \leq x \leq 0.132$. The two-gap superconductivity
typical of MgB$_2$ persists up to $x= 0.105$. In this region, the
values of the gaps $\Delta\ped{\sigma}$ and $\Delta\ped{\pi}$ were
determined by fitting the Andreev-reflection conductance curves
with a two-band Blonder-Tinkham-Klapwijk (BTK) model, and
confirmed by the single-band BTK fit of the $\sigma$- and
$\pi$-band conductances, separated by means of a magnetic field.
At $x=0.132$, when $T\ped{c}=19$~K, we clearly observed for the
first time the merging of the two gaps into one of amplitude
$\Delta\simeq 3$~meV.
\end{abstract}
\maketitle

In the last three years, the great experimental and theoretical
efforts of the scientific community have led to a clarification of
most features of the intermetallic superconductor MgB$_2$. These
features are mainly related to the presence of two band systems
($\sigma$ and $\pi$) and of the relevant gaps \cite{Liu,Brinkman}.
Point-contact spectroscopy (PCS) has proved particularly useful in
measuring both the $\sigma$- and $\pi$-band gaps at the same time
\cite{Szabo,Gonnelli} and determining the temperature dependency
of these gaps with great accuracy \cite{Gonnelli}. Very soon after
the discovery of superconductivity in MgB$_2$, substitutions of Al
for Mg and of C for B were tried, in order to introduce impurities
in the compound and modify its superconducting properties
\cite{Cava}. In particular, nearly single-phase
Mg(B$\ped{1-x}$C$\ped{x}$)$_2$ polycrystals with $0.09 \leq x \leq
0.13$ were obtained by starting from Mg and B$_4$C
\cite{Ribeiro,Avdeev}, which showed a linear dependence of the
cell parameter $a$ on the C concentration \cite{Avdeev}. More
recently, C-substituted MgB$_2$ single crystals were grown and
many of their structural, superconducting and transport properties
were measured \cite{Lee,Kazakov}. The first STM and PCS
measurements on polycrystalline Mg(B$\ped{1-x}$C$\ped{x}$)$_2$
have shown the persistence of the two gaps up to $x=0.1$
\cite{Schmidt,Samuely,Holanova}. Up to now, the predicted
achievement of single-gap superconductivity at a very high
impurity level has never been observed.

This Letter presents the results of PCS measurements in
Mg(B$\ped{1-x}$C$\ped{x}$)$_2$ single crystals with $0.047 \leq x
\leq 0.132$, in the presence of magnetic fields up to 9 T either
parallel or perpendicular to the \emph{c} axis. These measurements
gave us the dependence of the two gaps ($\Delta\ped{\pi}$ and
$\Delta\ped{\sigma}$) on the carbon content $x$ and showed that,
up to $x\simeq 0.10$, the two-gap superconductivity typical of
unsubstituted MgB$_2$ is retained. At $x=0.132$, we clearly and
reproducibly observed for the first time the merging of
$\Delta\ped{\pi}$ and $\Delta\ped{\sigma}$ into a single gap
$\Delta = 3.2 \pm 0.9$ meV which shows a ratio $2\Delta / k\ped{B}
T\ped{c}$ very close to the standard BCS value.

The high-quality Mg(B$\ped{1-x}$C$\ped{x}$)$_2$ single crystals were
grown at ETH (Zurich) with the same high-pressure technique adopted
for unsubstituted MgB$_2$ \cite{Kazakov}, and by using either
graphite powder or silicon carbide as a carbon source. Details on
the structural and superconducting properties of these crystals can
be found in a recent paper \cite{Kazakov}. The 7 different carbon
contents of our crystals were estimated from the lattice parameter
\emph{a}, assuming its linear dependence on $x$ \cite{Avdeev}. The
resulting $x$ values range between $x=0.047$ and $x=0.132$,
corresponding to bulk critical temperatures between 35~K and 19~K.

We performed PCS measurements with the current mainly injected along
the \emph{ab}-planes of the crystal, since in unsubstituted MgB$_2$
this is the most favourable configuration for the contemporaneous
measurement of both gaps \cite{Brinkman,Gonnelli}. The point
contacts were thus made on the flat side surface of the crystals
(not thicker than 80 $\mu$m) by using a small ($\varnothing \lesssim
50\,\mu$m) spot of Ag conductive paint. This ``soft'' version of the
PCS technique \cite{Gonnelli,Gonnellib} yields greater contact
stability on thermal cycling and greater reproducibility of the
conductance curves. By applying short current or voltage pulses to
the junctions, we were able to tune their characteristics and
achieve in most cases a normal-state resistance between 50 and
300~$\Omega$.
\begin{figure}[t]
\begin{center}
\includegraphics[keepaspectratio, width=0.7\columnwidth]{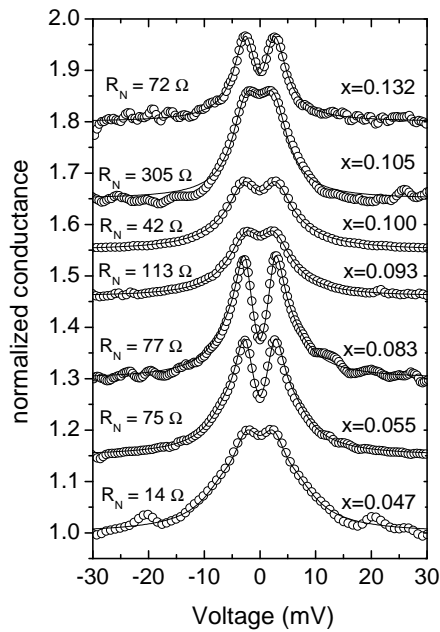}
\end{center}
\vspace{-5mm} %
\caption{\small{Normalized conductance curves at 4.2~K of
different \emph{ab}-plane junctions in
Mg(B$\ped{1-x}$C$\ped{x}$)$_2$ crystals with $0.047 \leq x \leq
0.132$ (open circles) and their two-band or single-band BTK fits
(solid lines). The curves are vertically shifted for clarity. The
best-fitting values of the parameters are shown in
Table~\ref{table:1}.}}\label{fig:1} 
\end{figure}
Since the in-plane mean free path in these single crystals ranges
from $\ell \simeq 17.5$~nm to $\ell \simeq 13$~nm for $x$ between
0.05 and 0.095 \cite{Kazakov}, these junctions result in the
ballistic regime. The formation of parallel micro-junctions
explains the very few cases in which ballistic conduction is
observed in low-resistance contacts.

Fig.~\ref{fig:1} reports some experimental conductance curves
($\mathrm{d}I/\mathrm{d}V$ vs. $V$) of point contacts on crystals
with different C contents (symbols). The curves are normalized as
explained in Ref.~\cite{Gonnelli}. As already shown by PCS in
Mg(B$\ped{1-x}$C$\ped{x}$)$_2$ polycrystals \cite{Holanova}, when $x
\geq 0.047$ the experimental curves do not show the clear four-peak
structure typical of \emph{ab}-plane contacts on unsubstituted
MgB$_2$ \cite{Gonnelli}. Hence, the proof of the presence of the
$\sigma$-band gap and its determination require a fit with the
two-band BTK model \cite{Szabo,Gonnelli,Samuely,Holanova,Gonnellib}
and/or the selective suppression of the $\pi$-band contribution to
the conductance, e.g. by applying a suitable magnetic
field~\cite{Gonnelli,Gonnellib}. In the following we will present
and discuss both these approaches.

First of all, let us discuss the fit of the zero-field conductance
curves reported in Fig.~\ref{fig:1}. In the two-band BTK model the
normalized conductance of a point contact is given by
$\sigma=(1-w\ped{\pi})\sigma\ped{\sigma}+w\ped{\pi}\sigma\ped{\pi}$
where $\sigma\ped{\sigma}$ and $\sigma\ped{\pi}$ are the partial
$\sigma$- and $\pi$-band conductances, respectively, and
$w\ped{\pi}$ is the weight of the $\pi$-band contribution. Thus, the
total number of parameters in the model is 7. In unsubstituted
MgB$_2$, $w\ped{\pi}$ ranges from 0.66 to 0.99, as
predicted~\cite{Brinkman} and confirmed by directional
PCS~\cite{Gonnelli}. In the absence of a similar prediction for the
C-substituted compound, we conservatively took $w\ped{\pi}$ between
0.66 and 0.8.

For any $x$ between 0.047 and 0.105, the two-band BTK model fits
very well the experimental data, as shown in Fig.\ref{fig:1} (solid
lines). The best-fitting parameters are listed in
Table~\ref{table:1}.
\begin{table}[b]
\vspace{-3mm}
  \centering
\begin{tabular}{|c|c|c|c|c|c|c|c|}
  \hline
  $x$                 & \,0.047\, & \, 0.055\, & \,0.083\, & \,0.093\, & \,0.100\, & \,0.105\, &  \,0.132\, \\
  \hline
  $\Delta\ped{\sigma}$ & 7.0   & 6.6   &   5.8 &  4.3  & 4.9   & 4.25  & 2.8 \\
  \hline
  $\Delta\ped{\pi}$   & 3.2   & 3.0   &  3.0 &  2.8  & 3.3  & 3.2  &  -- \\
  \hline
  $\Gamma\ped{\sigma}$ & 3.15 & 2.50 & 2.60 & 3.20 & 4.55 & 2.55 & 1.50\\
  \hline
  $\Gamma\ped{\pi}$ & 1.60 & 1.10 & 0.91 & 2.00 & 2.07 & 1.62 & -- \\
  \hline
  $w\ped{\pi}$ & 0.66 & 0.75 & 0.70 & 0.70 & 0.69 & 0.80 & 0 \\
  \hline
\end{tabular}
  \caption{\small{Values of the gaps and of the broadening parameters
  for the best-fit curves of Fig.\ref{fig:1} (solid lines). $Z\ped{\sigma}$
  is always close to 0.50 and $Z\ped{\pi}$ ranges between 0.34 and 0.57.}} \label{table:1}
\end{table}
In the crystals with the highest C content, i.e. $x=0.132$, the
two-band BTK fit requires gap values very close to each other and,
practically, interchangeable (in the sense that their error bars
largely overlap). Actually, a fit with the standard
\emph{single-band} BTK model works even better. The solid line
superimposed to the conductance curve at $x=0.132$ in
Fig.\ref{fig:1} is indeed obtained with only one gap of amplitude
$\Delta = 2.8 \pm 0.2$~meV.

The reliability of the determination of $\Delta\ped{\sigma}$ and
$\Delta\ped{\pi}$ by means of a 7-parameter fit may be questioned.
We thus tried to apply here the procedure we already used in
unsubstituted MgB$_2$ \cite{Gonnelli}, that consists in separating
the partial $\sigma$ and $\pi$-band contributions to the total
conductance ($\sigma\ped{\sigma}$ and $\sigma\ped{\pi}$) by means
of a suitable magnetic field, and in fitting each of them with the
standard, three-parameter BTK model. A detailed discussion of the
applicability of the BTK fit of Andreev reflection curves in the
presence of magnetic fields is given in
Ref.~\onlinecite{Gonnellib}.
\begin{figure}[t]
\begin{center}
\includegraphics[keepaspectratio, width=0.7\columnwidth]{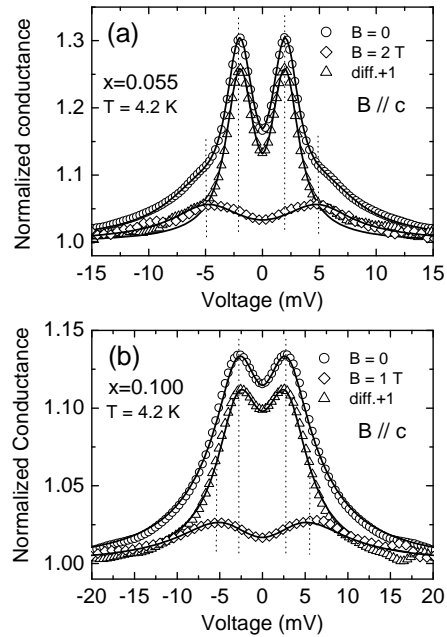}
\end{center} \vspace{-5mm} \caption{\small{(a) Normalized conductance
curves of a $ab$-plane contact in a crystal with $x=0.055$, at
$B$=0 (circles) and $B$=2~T (diamonds). Triangles represent the
difference of the previous two curves, shifted by 1. Solid lines:
best-fitting curves given by the two-band (upper curve) or
single-band (lower curves) BTK model. (b): same as in (a), but for
a crystal with $x=0.100$. Here the applied magnetic field is
$B$=1~T. Vertical lines indicate the conductance
peaks.}}\label{fig:2}
\end{figure}
\begin{figure}[t]
\begin{center}
\includegraphics[width=0.7\columnwidth]{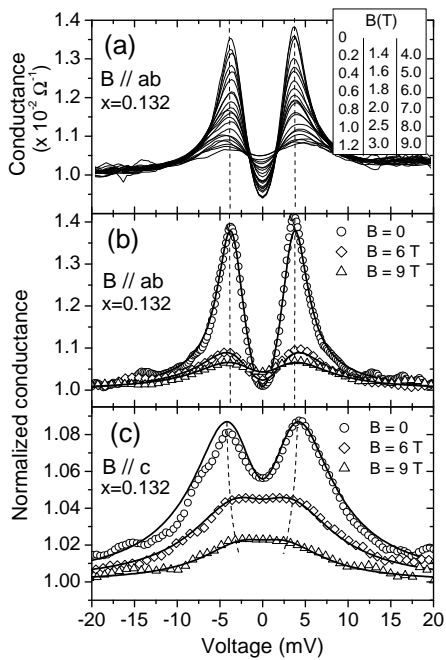}
\end{center}
\vspace{-5mm} \caption{\small{(a) Raw conductance curves at
T=4.2~K in a crystal with $x=0.132$, in a magnetic field
$\mathbf{B}\parallel ab$. (b) Some of these curves after
normalization (symbols). Solid lines are the single-band BTK fits.
On increasing the field from 0 to 9 T, the best-fit parameters
vary as follows: $\Delta$=4.0$\rightarrow$3.5~meV, $\Gamma$=0.8$
\rightarrow$3.8~meV, $Z$=0.60$ \rightarrow $0.53. (c) Same as in
(b), but with $\mathbf{B}\parallel c$. Here, the parameters vary
as follows: $\Delta$=4.0$ \rightarrow$2.0~meV, $\Gamma$=3.6$
\rightarrow $4.6~meV, $Z$=0.52$ \rightarrow$0.42. In (a), (b) and
(c), dashed lines indicate the conductance peaks.
}}\label{fig:3}
\end{figure}

Fig.~\ref{fig:2} shows how this works in crystals with $x$=0.055
(a) and $x$=0.100 (b). In both panels, the zero-field conductance
$\sigma\ped{0}$ (circles) is compared to the relevant two-band BTK
fit (solid line). Diamonds represent instead the conductance
$\sigma\ped{B^*}$ measured in a magnetic field $B^*$ (making an
angle $\varphi=90\pm 2^{\circ}$ with the $ab$ planes) that
completely removes any structure related to the $\pi$-band gap
\footnote{Similar results were obtained for $\mathbf{B}\parallel
ab$, as we will discuss in a forthcoming paper.}. For $x$=0.055,
$B^{*}\!\simeq\! 2$~T, while for $x$=0.100 $B^{*}\!=\!1$~T, as in
unsubstituted MgB$_2$. Incidentally, this indicates that $B^{*}$
has a maximum somewhere between $x=0$ and $x=0.100$, like the
critical field \cite{Masui,Wilke} and the irreversibility field
\cite{Ohmichi}. $\sigma\ped{B^{*}}$ contains only the
$\sigma$-band contribution to the conductance and can thus be
fitted by taking $\sigma\ped{\pi}=1$ in the two-band BTK model.
Since $w\ped{\pi}$ is reasonably field-independent, only
$\Delta\ped{\sigma}$, $\Gamma\ped{\sigma}$ and $Z\ped{\sigma}$
remain as adjustable parameters. Finally, the difference
$\sigma\ped{diff}=\sigma\ped{0}-\sigma\ped{B^{*}}+1$ (triangles)
contains only the $\pi$-band conductance and can thus be fitted by
taking $\sigma\ped{\sigma}=1$ in the two-band BTK model
\cite{Gonnelli}.

\begin{figure}[t]
\begin{center}
\includegraphics[width=0.7\columnwidth]{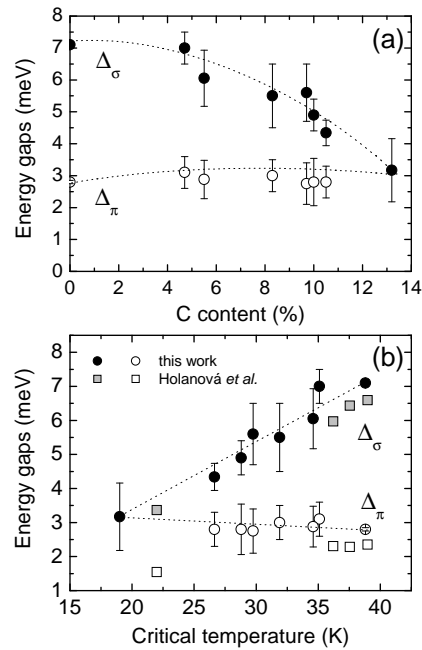}
\end{center}
\vspace{-5mm} \caption{\small{Gap amplitudes, $\Delta\ped{\sigma}$
and $\Delta\ped{\pi}$, as a function of the C content $x$ (a) and
of the bulk critical temperature $T\ped{c}$ (b). In the latter
case, data from PCS in polycrystals \cite{Holanova} are reported
for comparison. Dotted lines are guides to the
eye.}}\label{fig:5}\vspace{-2mm}
\end{figure}

The separate fit of $\sigma\ped{\sigma}$ and $\sigma\ped{\pi}$
gives the following results: (a) $\Delta\ped{\sigma}$=5.45 meV,
$Z\ped{\sigma}$=0.475, $\Gamma\ped{\sigma}$=2.4 meV and
$\Delta\ped{\pi}$=2.30 meV, $Z\ped{\pi}$=0.488,
$\Gamma\ped{\pi}$=0.485 meV; (b) $\Delta\ped{\sigma}$=4.9 meV,
$Z\ped{\sigma}$=0.525, $\Gamma\ped{\sigma}$=4.55 meV and
$\Delta\ped{\pi}$=3.28 meV, $Z\ped{\pi}$=0.42,
$\Gamma\ped{\pi}$=2.07 meV. In the case of $x=0.055$, a slight
reduction in $\Delta\ped{\sigma}$ (smaller than the gap
distribution width, see Fig.~\ref{fig:5}) is present with respect
to the two-band fit (that gave $\Delta\ped{\sigma}$=6.05 meV and
$\Delta\ped{\pi}$=2.35 meV), possibly because $B^{*}$=2~T is
already comparable to $B\ped{c2}^{\parallel c}\!\!\simeq$8~T
\cite{Kazakov}. For the case of $x=0.100$, the parameters coincide
with those reported in Table~\ref{table:1}. Similar agreement was
found for any C content up to $x=0.105$, and in all the junctions
we studied, showing that this procedure has a high level of
internal consistency and gives precise and reliable results, as in
unsubstituted MgB$_2$ \cite{Gonnelli}.

\begin{figure}[t]
\begin{center}
\includegraphics[keepaspectratio, width=0.9\columnwidth]{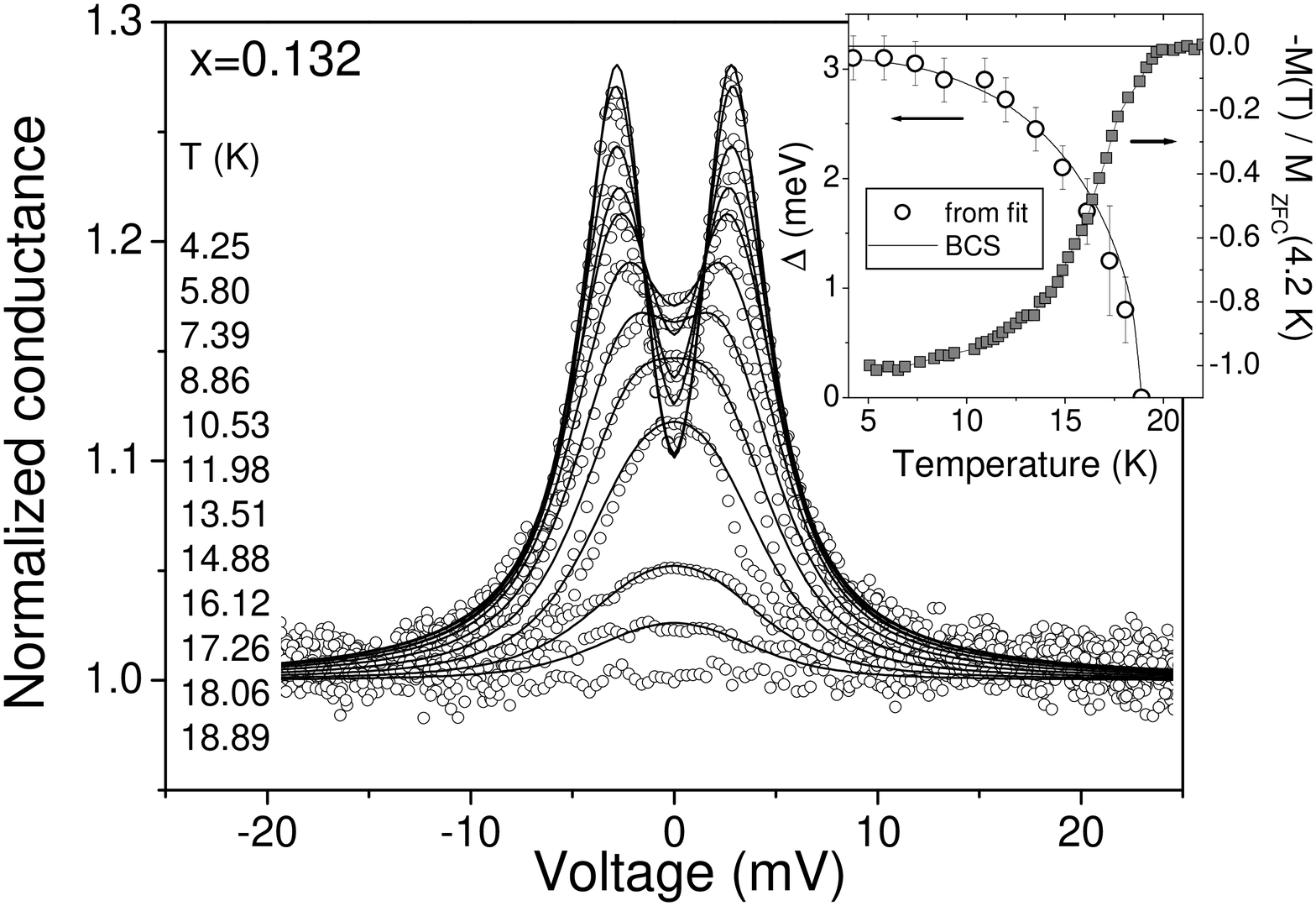}
\end{center}
\vspace{-6mm} %
\caption{\small{An example of temperature dependence of the
normalized conductance curves for $x$=0.132 (symbols) with the
relevant fit (solid lines). The gap given by the fit (inset, open
symbols) follows very well the BCS $\Delta(T)$ curve (inset,
line). Filled symbols in the inset represent the ZFC
magnetization.}}\label{fig:4}
\end{figure}

In the crystals with $x=0.132$, the same procedure gives quite
different results and further confirms the presence of a single
gap. Fig.~\ref{fig:3}(a) reports an example of magnetic-field
dependence of the raw conductance curves, for $\mathbf{B}\parallel
ab$. Contrary to what happens at lower C contents, there is no
clear shift of the conductance maxima towards higher energy (which
is the hallmark of the suppression of the $\pi$-band contribution
at $B=B^{*}$) at any field between 0 and 9~T. Rather, the peaks
approximately remain in the same position, as indicated by
vertical dashed lines. In Fig.~\ref{fig:3}(b) the curves at B=0
(circles), B=6~T (diamonds) and B=9~T (triangles) are shown after
normalization. Similarly, Fig.~\ref{fig:3}(c) reports the
normalized conductance curves of another contact, measured at the
same fields but in the $\mathbf{B}\parallel c$ configuration. In
both (b) and (c), the experimental curves are compared to their
single-band BTK fit, whose parameters are reported in the caption.
The good quality of the fit and the magnetic-field dependence of
the conductance curves strongly indicate that, at $x=0.132$,
Mg(B$\ped{1-x}$C$\ped{x}$)$_2$ is an anisotropic single-gap
superconductor with $B\ped{c2}^{\parallel
ab}\!>\!B\ped{c2}^{\parallel c}\!>\!9$~T.

Further support to the presence of a single gap in crystals with
$x=0.132$ comes from the temperature dependence of the conductance
curves, shown in Fig.\ref{fig:4}. All the experimental curves
(symbols) can be fitted by the single-band BTK model (solid
lines), and the resulting gap values (inset, open circles) follow
very well the BCS curve with $2\Delta/k\ped{B}T\ped{c}=3.8$
(inset, line). A comparison of $\Delta(T)$ with the ZFC
magnetization (inset, filled squares) shows that the critical
temperature of the junction coincides with the bulk $T\ped{c}$ and
that the magnetic transition is complete at 5~K.

The dependence of the gaps on the carbon content and on the bulk
$T\ped{c}$ is reported in Fig.~\ref{fig:5} (a) and (b),
respectively. Each point results from an average of various gap
values (usually 4-8) obtained in different contacts. Hence, error
bars indicate the maximum spread of measured values, and give an
idea of the good reproducibility of our results. The value of the
single gap at $x=0.132$, $\Delta=3.2 \pm 0.9$~meV, and the bulk
$T\ped{c}=19$~K give a gap ratio $2 \Delta/k\ped{B}T\ped{c}\sim
3.9$ close to the BCS value. Despite the large uncertainty on
$\Delta$, all the curves at $x=0.132$ were best-fitted by a single
gap BTK model.
Note that the gap merging at $x=0.132$ is \emph{perfectly
consistent} with the regular and smooth trend of the gaps for
lower C contents. The decrease in $\Delta\ped{\sigma}$ and the
slight increase in $\Delta\ped{\pi}$ shown in Fig.~\ref{fig:5}
suggest an increase in \emph{interband} scattering, as predicted
by the two-band model \cite{Liu}. However, interband scattering
alone would make the gaps merge in a isotropic gap $\Delta\simeq
4$~meV when $T\ped{c}$=26~K. Clearly, other effects are playing a
role, such as the changes in the electronic structure due to
electron doping \cite{Singh} and the hardening and narrowing of
the $E\ped{2g}$ phonon mode \cite{Masui}. By taking into account
these effects, the observed $\Delta\ped{\sigma}(x)$ and
$\Delta\ped{\pi}(x)$ curves in Mg(B$\ped{1-x}$C$\ped{x}$)$_2$
single crystals (included their merging at $T\ped{c}=19$~K) can be
well explained within the two-band Eliashberg theory as resulting
from the band filling, plus a decrease in the Coulomb
pseudopotential and an \emph{increase} in interband scattering
\cite{Kortus,UmmarinoC} -- even though this contrasts with the
theoretical prediction that C substitution should have negligible
effects on the $\sigma - \pi$ scattering \cite{Erwin}.

In Fig.~\ref{fig:5}(b), our gap values are compared to data from PCS
in polycrystals by Ho\v{l}anov\'{a} et al. \cite{Holanova}. Apart
from a small systematic shift, their $\Delta\ped{\sigma}(T)$ curve
is in good agreement with our results in the whole doping range,
while their value of $\Delta\ped{\pi}$ at $x=0.10$
($T\ped{c}=22$~K), is much smaller than ours. This disagreement is
probably related to the greater amount of interband scattering in
our crystals, which might be due to microscopic defects (also acting
as pinning centers \cite{Kazakov}) or to the existence of
micro-domains with ordered C distribution \cite{UmmarinoC}. Both
possibilities are compatible with the presence of local C-content
inhomogeneities on a scale comparable to $\xi$ \cite{Kazakov}.
Finally, notice that the evidence of single-gap superconductivity at
$x=0.132$, accompanied by an anisotropic bandstructure (see
Fig.\ref{fig:3}) is consistent with the extrapolation at $x>0.10$ of
recent measurements of $B\ped{c2}$ and Hall effect in single
crystals up to $x=0.10$ \cite{Masui}.

In conclusion, we have presented the results of the first
directional point-contact measurements in Mg(B$_{1-x}$C$_x$)$_2$
single crystals with $0.047 \le x \le 0.132$, that allowed us to
obtain the dependence of $\Delta_{\sigma}$ and $\Delta_{\pi}$ on
the carbon content $x$. This dependence was confirmed by applying
to the junctions a suitable magnetic field $B^*$ able to remove
the contribution of the $\pi$-band gap to the total conductance,
thus allowing the separate determination of the gaps via a
single-band BTK fit. Up to $x \sim 0.10$, the two-gap nature of
superconductivity characteristic of unsubstituted MgB$_2$ is
retained. At $x=0.132$ we clearly and reproducibly observed for
the first time the merging of the two gaps into a single gap
$\Delta \simeq 3$ meV with a gap ratio $2\Delta/k_BT_c$ close to
the standard BCS value and a critical field greater than 9 T.

This work was done within the INFM Project PRA ``UMBRA'', the
INTAS Project n.01-0617 and the FIRB Project RBAU01FZ2P. V.A.S.
acknowledges the support of RFBR Projects n. 04-02-17286 and
02-02-17133, and of the Russian Ministry of Science and Technology
(Contract n. 40.012.1.1.1357). The work in Zurich was supported by
the Swiss National Science Foundation, the NCCR program MaNEP and
ETH. 
\end{document}